\documentclass[12pt, letterpaper]{article}

\usepackage{amsmath}
\usepackage{amsfonts}
\usepackage{amssymb}
\usepackage{graphicx}

\usepackage{color}

\usepackage{amsmath, amssymb, graphics, epsfig, graphicx}
\usepackage{epsf}
\usepackage{epstopdf}
\usepackage {amssymb}
\newcommand{\nc}{\newcommand}
\nc{\ba}{\begin{eqnarray}}
\nc{\ea}{\end{eqnarray}}
\newcommand\be{\begin{equation}}
\newcommand\ee{\end{equation}}

\newcommand{\bea}{\begin{eqnarray}}
\newcommand{\eea}{\end{eqnarray}}

\begin{document}

\vspace{5mm}
\vspace{0.5cm}
\begin{center}

\def\thefootnote{\fnsymbol{footnote}}

{\Large   The spectrum  of perturbations  inside the Schwarzschild  black hole }
\\[0.5cm]

{ Hassan Firouzjahi}

{\small \textit{School of Astronomy, Institute for Research in Fundamental Sciences (IPM) \\ P.~O.~Box 19395-5531, Tehran, Iran }}\\

\vspace{0.1cm}
{\small {e-mail:    firouz@ipm.ir }}\\

\end{center}

\vspace{.8cm}

\hrule \vspace{0.3cm}


\begin{abstract}
We study the spectrum of the bound state perturbations in the interior of the Schwarzschild black hole for the scalar, electromagnetic  and gravitational perturbations.  Demanding that the perturbations to be regular at the center of the black hole   determines  the spectrum of the bound state solutions. We show that our analytic expression for the spectrum is in very good agreement with the imaginary parts of the high overtone 
quasi normal mode excitations obtained for the exterior region. We also present a simple scheme to  calculate the spectrum  numerically to good accuracies. 


\end{abstract}
\vspace{0.5cm} \hrule
\def\thefootnote{\arabic{footnote}}
\setcounter{footnote}{0}
\newpage
\section{Introduction}

 Black holes have played important roles in the developments of theoretical physics. During the past decades  astronomers have suggested the existence of massive and supermassive black holes in the center of typical galaxies.  In addition, the recent detection of gravitational waves by the LIGO team \cite{Abbott:2016blz, Abbott:2016nmj} 
from the merging of binary astrophysical black holes have left no doubt on the reality of black holes in Nature.  On the theoretical side black hole physics plays a key role in understanding of quantum gravity which was studied extensively  in the past decades. 

The perturbations of the black hole were first studied by Regge and Wheeler \cite{Regge:1957td}, for a review see
\cite{Chandrasekhar:1985kt}.  These perturbations are the characteristics of black hole in its final stage of oscillations, encoding information about its mass, angular momentum and charge. In this view, the oscillations of black hole in its final ringing stage can be used as a tool in gravitational wave astronomy observations, such as LIGO, to determine its fundamental parameters. There have been numerous studies of black hole perturbations for its quasi normal mode (QNM) excitations. These are defined as the outgoing perturbations for  region far from the center of black hole and ingoing waves on the surface of horizon. The conventional view is that once the ingoing QNMs cross the horizon and enter the black hole, they escape the observations and are not accessible to the external observer. 
This view  leaves the final fate of the ingoing perturbations unanswered.


The physical processes governing the dynamics of the  interior of  the black hole are not well understood. The prime reason is that  the interior  region of black hole is causally disconnected from the exterior region. As mentioned, any in-falling signal smoothly passes through the event horizon and no signal can escape from inside the event horizon. In addition, inside the 
black hole the roles of coordinates $ t$ and $r$ as the timelike and spacelike coordinates are switched. In this view, the interior of the black hole is like a cosmological background bounded by event horizon.  The global structure of spacetime suggests that the future singularity $r=0$ behaves as the onset of cosmological big crunch  singularity. Any signal generated from 
the past  null infinity ${\cal I}^{-}$  propagates towards the future  singularity $r=0$. The black hole spacetime can be viewed as an anisotropic cosmological background with the topology $R\times S^2$ known as the Kantowski-Sachs \cite{Kantowski:1966te} spacetime.  On the other hand, perturbations in cosmological backgrounds are studied extensively. Indeed, it is believed that all structure in observable Universe are generated from tiny quantum fluctuations generated during primordial inflation. It is therefore a natural question to study perturbations  inside the black hole as a particular cosmological background.  The main goal of this paper is to study the perturbations inside the black hole. 

As we shall see below, the effective potential for the perturbations in the interior region can support the bound state solutions in which the spectrum is pure imaginary i.e.  $\omega = i \Gamma$ with real values of $\Gamma$.  This is unlike the spectra of QNMs in which the specific boundary conditions requires that $\omega = \omega_R - i \Gamma$ with $\Gamma >0$.

 \section{Quasi normal mode perturbations}
 \label{QSN-sec}
 
To set the stage for studying the evolution of perturbations for the interior of black hole, here we quickly review the standard results for the QNM perturbations of the Schwarzschild black hole in its exterior region. 

Working with the convention $G=1$ and setting the black hole mass to unity, $M=1$, the 
 Schwarzschild metric  is given by
\ba
\label{metric-Sch}
ds^2 = - (1- \frac{2}{r} ) dt^2 + \frac{d r^2}{(1- \frac{2}{r} )} + r^2 d \Omega^2 \, ,
\ea
in which $t$ is the time, $r$ is the radial coordinate and $d \Omega^2 $ represents the angular components of the metric. 

We study the axial metric perturbations, the massless scalar field and the electromagnetic field perturbations  which are 
collectively denoted by $Z$. We leave aside the polar perturbations due to the complexity of its potential but our results apply for polar perturbations as well. Note that the field $Z$ is the canonically normalized field such that the equations of the perturbations take the form of Schrodinger equation given by the Regge-Wheeler equation \cite{Regge:1957td}
\ba
\label{eq-tort}
  \frac{d^2 Z}{d r_*^2}  + \left(\omega^2   -   V_{eff} \right) Z =0  \, ,
\ea
in which $r_*$ is the tortoise coordinate defined via $dr_* = \frac{r \, d r}{(r - 2 )}$.  

The effective potential is given by \cite{Chandrasekhar:1985kt, Nollert:1999ji}
\ba
\label{Veff}
V_{eff} = \left(1- \frac{2}{r} \right)  \left[ \frac{2 \sigma}{r^3} + \frac{\ell (\ell+1)}{r^2} \right ] \, ,
\ea
in which $\ell$ is the usual angular momentum and 
the parameter $\sigma$ takes the value $1, 0, -3$ for the scalar, electromagnetic and  gravitational perturbations respectively. More specifically, one can set $\sigma =  (1-s^2)$ for $s$ being the spin of the corresponding field, $s=0, 1, 2$  for the scalar, electromagnetic and gravitational fields respectively.

For the exterior of the black hole, $r>2$, we have 
\ba
\label{r-star-ext}
r_* = r + 2 \ln(\frac{r}{2} -1)  \,,        \quad \quad (r>2)  \, ,
\ea 
so 
$-\infty < r_* < +\infty$, in which the horizon $r=2$ is mapped to $r_*=-\infty$. The QNM perturbations are defined as the outgoing wave for $r_*\rightarrow +\infty$,  $ Z \sim e^{-i \omega (t- r_*) }$,  and ingoing solution for $r_*\rightarrow -\infty$, $ Z \sim e^{-i \omega (t+ r_*) }$ \cite{Chandrasekhar:1975zza}. 
The spectrum of the quasi normal modes are studied extensively in the literature, for a review see \cite{ Nollert:1999ji, Kokkotas:1999bd, Berti:2009kk} and the references therein. 

A natural question which comes in mind is what happens to the perturbations in the interior of the black hole. In the interior of the black hole the roles of $(t, r)$ as the timelike and spacelike coordinates are switched and indeed the spacetime represents an anisotropic cosmological background with $r=0$ becoming a  future cosmological singularity, i.e. a big crunch. 

In terms of their origins, the perturbations inside the black hole can be divided in two different categories. The first category contains perturbations which are generated in the exterior region and cross the horizon and hit the singularity at $r=0$.  These perturbations  are basically the extension of the QNM solution to the interior region. To construct these solutions in the interior region, one has to impose the continuity of the solution across the horizon, expressing the solution in terms of the  Kruskal-Szekeres coordinate which are continuous across the horizon.  The second category contains perturbations which are bounded in the interior of the black hole and are causally disconnected from the exterior observer. As we shall see below, the structure of the effective potential for the interior region supports the existence of this bound state solution.
In this work, we study the bound state solution, i.e. perturbations which are bounded to the interior of the black hole. However, a possible link between the spectrum of the bound states and the incoming QNMs will be speculated in next Section.

\section{Perturbations inside the black hole}
\label{inside-pert}

Now we concentrate on the perturbations inside the black hole, $r<2$. As in the exterior region, the symmetry of the configuration suggests the profile $Z= e^{i \omega t} f(r)$. However, we should keep in mind that $t$ is a spacelike coordinate with no limitation on its range  so we allow $-\infty < t < +\infty$. With this picture in mind,  the form of the perturbation equation is formally the same as Eq. (\ref{eq-tort}) but $r_*$ now is given by 

\ba
\label{r-star-int}
r_* = r + 2 \ln(1-\frac{r}{2} )  \,,        \quad \quad (r<2)  \, ,
\ea
in which  $-\infty< r_* < 0$, with horizon being mapped to $r_*=-\infty $ while the singularity is located at 
$r_* =0$. 
The logarithmic branch cut in the two expressions of $r_*$ given in Eqs. (\ref{r-star-ext}) and (\ref{r-star-int}) is evident.

\begin{figure}[t]
\begin{center}
	\includegraphics[scale=0.5]{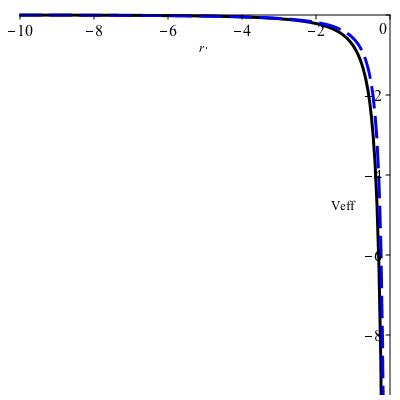} 
	\hspace{1.5cm}
	\includegraphics[scale=0.5]{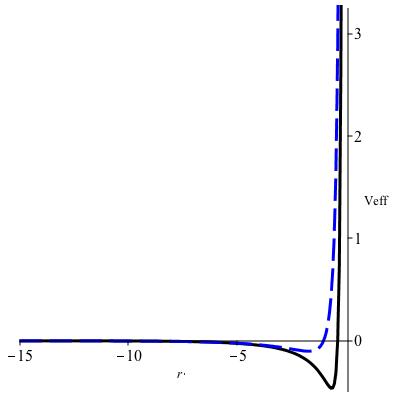}
	\end{center}
\caption{ Effective potential  as a function of $r_*$. Left: scalar perturbations ($\sigma=1$) with $\ell=1$; right: gravitational perturbations ($\sigma=-3$) with $\ell=2$. In both plots, the solid black curves are the full potential  Eq. (\ref{Veff}) while the dashed blue curves are our approximate potential Eq. (\ref{V-app})}
\label{V-plot}
\end{figure}

While the form of the equation is the same as in the exterior region, however its interpretation is quite different. With $r_*$ being a timelike coordinate for the interior region, Eq. (\ref{eq-tort}) is a time dependent differential equation. This is similar 
to the  equation for scalar perturbation  in inflationary backgrounds, the Sasaki-Mukhanov equation, with
$\omega^2$ playing the role of the corresponding comoving wave number.  
In addition, the shape of effective potential is very different for the interior region. In Figure \ref{V-plot} the shapes of the potential for the scalar and gravitational perturbations 
are presented. For the scalar perturbations, we see that the potential is monotonically decreasing. This should be compared with the effective potential for the exterior region in which the effective potential has a bump. For the gravitational perturbations, the situation is somewhat different in which the potential has a negative global minimum and reaches to 
$+\infty$ near the singularity.  

The shape of the potential suggests that the bound states with $\omega^2 <0$ exist. This is similar to negative energy in the effective Schrodinger equation in the central force problem such as in Hydrogen atom. For these bound state solutions 
the perturbations fall off exponentially towards $r_* \rightarrow -\infty$ so they can not leak out of the black hole. This is in line with the interpretation that the interior of the black hole is causally disconnected from the exterior region, so any perturbation generated inside the black hole is inaccessible to the exterior observer. With $\omega^2 <0$ we have a pure imaginary $\omega$, suggesting that the profile containing $e^{i\omega t}$ diverges either at $t=-\infty$ or $t=+\infty$, depending on the sign of $\omega$. This may raise concern that the solution has a diverging spatial profile. However, we should recall that in the interior region Eq. (\ref{eq-tort}) is not a propagating wave equation  to enforce us taking $t$ moving towards $+\infty$. Specifically, one can  consider the slices of constant $t$ and let the system evolves in time from $r_*=-\infty$ to $r_*=0$.

Because of the non-trivial form of the effective potential, it is not possible  to find exact analytic solution for the 
wavefunction and the spectrum of perturbations. This is in par with the QNM analysis in the exterior region in which it was not possible so far to find exact analytic solution.   However, several approximation approaches were employed to  obtain useful analytic results for QNMs perturbations,  such as in using the bound state of the inverted potential  \cite{ Ferrari:1984ozr, Ferrari:1984zz} or the WKB approximations \cite{Schutz:1985zz, Iyer:1986np}. Similarly, here we try to find some analytic expressions for the perturbations inside the black hole using appropriate approximations. 

First, we find the asymptotic form of the effective potential at both of its tails. Near the singularity $r=0$, we have $r_* \simeq -r^2/4 $ while near the horizon $r=2$ we have $1-\frac{2}{r} \rightarrow -e^{(r_*/2)-1}$.
Consequently, the effective potential has the following asymptotic forms
\ba
\label{V-asym}
V_{eff} (r_* \rightarrow 0) \rightarrow -\frac{\sigma}{4 r_*^2} ,  \quad  \quad
V_{eff} (r_* \rightarrow -\infty) \rightarrow -\frac{\ell (\ell +1) +\sigma}{4 e}\,  e^{\frac{r_*}{2}} \, .
\ea

With the above asymptotic behaviours, we approximate the effective potential with the following potential
\ba
\label{V-app}
V_{eff} = - \frac{ \alpha}{64 \sinh^2(\frac{r_*}{4})} - \frac{\beta}{64} \frac{1+ \tanh(\frac{r_*}{4})}{ \tanh(\frac{r_*}{4})^2} \, .
\ea
Here the parameters $\alpha$ and $\beta$ are chosen such that the above approximate potential coincides with the asymptotic forms of the exact effective potential:
\ba
\alpha = 3 2 L - \sigma , \quad \quad \beta = 2 \sigma - 32 L   \,  , 
\ea
in which we have defined 
\ba
L \equiv \frac{\ell (\ell +1) +\sigma}{4 e} \, .
\ea
In  Figure \ref{V-plot} the approximate effective potential Eq. (\ref{V-app}) is compared with the full effective potential. As expected, the two potentials match with each other on their tails while they deviate in the interval to some extent. However, we expect the spectrum of the perturbations, which are mainly determined from the boundary conditions at the center and at
$r_* =-\infty$, to be closely captured by our approximate effective potential.

Happily the perturbation  equation associated with the approximate potential  Eq. (\ref{V-app}) can be solved exactly. We relegate the details of the analysis of the bound state solution to Appendix \ref{app1} and here present the main results. The profile of the wavefunction is obtained to be 
\ba
\label{sol-final}
Z = C \Big( 1+ \tanh(\frac{r_*}{4}) \Big)^{\pm2 i \omega } \Big( 1- \tanh(\frac{r_*}{4}) \Big)^{\mp2 i \omega-\frac{1}{2} -\frac{\sqrt{1-\sigma}}{2} }  \Big( \tanh(\frac{r_*}{4}) \Big)^{\frac{1}{2} +\frac{\sqrt{1-\sigma}}{2}}  \nonumber\\
 \times \, {_2}F_1\left( a, b,c;   \frac{1+ \tanh(\frac{r_*}{4})}{ 1- \tanh(\frac{r_*}{4})}  \right) \, ,
\ea
in which ${_2}F_1$ is the hypergeometric function  with the parameters  $a, b$ and $c$ given by 
\ba
a& =& \frac{1}{2} +  \sqrt{- 4 \omega^2 -\frac{\beta}{8}}  + \frac{\sqrt{1- \sigma}}{2} \pm 2 i \omega \, ,   \nonumber\\
b &= &  \frac{1}{2} -\sqrt{- 4 \omega^2 -\frac{\beta}{8}}  + \frac{\sqrt{1- \sigma}}{2} \pm 2 i \omega  \, ,  \nonumber\\
c &= & 1\pm 4 i \omega  \, .
\ea

Now we impose the condition that the wave function to be regular at the center $r_*=0$. Physically, this corresponds to the requirement that the  perturbations do not modify the background solution, i.e. they do not change the singularity structure of the background.  If the wavefunction diverges at $r_*=0$, then one expects that the energy associated with these perturbations becomes infinite and the back-reactions of the perturbations on the background geometry can not be neglected. 

The asymptotic form of the wave function for $r_* \rightarrow 0$ is given by
\ba
Z \rightarrow r_*^{\frac{1}{2} +\frac{\sqrt{1-\sigma}}{2}}  {_2}F_1\left( a, b,c; 1+ \frac{r_*}{2} \right) \, .
\ea 
The condition that the wave function to be regular requires that the 
hypergeometric function to truncate to polynomials. This in turn requires that either $a$ or $b$ to be a non-positive integer, $-n, n\ge 0$. In either case, choosing the asymptotic form of the bound state to have the form $Z \propto e^{i \omega r_*}$, 
the spectrum of the perturbation is obtained to be (see Appendix \ref{app1} for further details)
\ba
\label{spectrum}
\omega_{(n)} = - \frac{i}{4} \Big( n+ \frac{1}{2} +  \frac{ \sqrt{1-\sigma} }{2} - \frac{ \frac{1}{e}(\ell +\frac{1}{2})^2 +\frac{4 \sigma - e\sigma -1}{4 e}    }{n+ \frac{1}{2} +  \frac{ \sqrt{1-\sigma} }{2}} 
\Big) \, , \quad  n=0,1,2... \, .
\ea
The above is the spectrum of the bound state i.e. $\omega^2 <0$ so $\omega$ is pure imaginary as expected.  

A few generic conclusions can be drawn from the form of $\omega_{(n)}$ given above.   First, we see that for large $n$, $\omega_{(n)}$ goes like $(n+ \frac{1}{2})/4$. Second, the correction from the angular momentum  (neglecting the prefactor $1/e$) goes like $ -\frac{(\ell +\frac{1}{2})^2}{(n+\frac{1}{2})}$.  Third, for large $n$ the spectrum becomes independent 
of $\sigma$. Interestingly, these generic properties are shared by the high overtone QNM perturbations as well.


Motivated by the above discussions, it is instructive to compare our result for the spectrum of the bound state with the  imaginary part of the QNMs for the exterior perturbations.  Comparing our result Eq. (\ref{spectrum}) with the numerical values of the black hole QNM  \cite{Leaver:1985ax, Nollert:1993zz, Liu},  we find that they are in excellent agreement for $n \gg1$. For example, Nollert has suggested 
the  following empirical formula for the imaginary part of the $\ell=2$ gravitational QNM spectrum \cite{Nollert:1993zz, Liu}
\ba
\label{Nollert}
\omega_{(N)} = -\frac{i}{4} ( n+\frac{1}{2} ) + \frac{0.6859 \, i}{2}  (2n +1)^{-1/2} \, .
\ea 
The agreement between our analytic expression with  his empirical result Eq. (\ref{Nollert}) is very good. For example, for 
$n \sim 200$ the agreement is better than 0.5\%  and it improves for larger $n$.  

Using the analogy of QNM spectrum with the Coulomb problem, Liu and Mashhoon have found \cite{Liu-Mashhoon}
\ba
\label{spectrum-LM}
\omega_{(LM)} = - \frac{i}{4} \Big( n+ \frac{1}{2}  - \frac{ (\ell +\frac{1}{2})^2    }{n+ \frac{1}{2}} \Big) \, .
\ea
We see that our expression Eq. (\ref{spectrum}) is in very good agreement  with their formula for $n\gg1$. 

Despite the agreements between the spectrum of our bound state solution and the imaginary part of  QNMs, however, 
the physical connection between these two solutions is missing. The main puzzle is that we have defined the bound state as the solutions which fall off on the boundary so they are inaccessible to the outside observer. However, QNMs are 
propagating waves crossing the horizon from the exterior regions.  Besides the imaginary parts, the spectrum of QNMs have a real part which is the hallmark of a propagating wave. Nonetheless, despite their fundamental differences, there may exist deep physical reasons why our bound state spectrum coincides with the imaginary components of the QNMs spectrum for $n \gg 1$.

\section{Numerical solutions}
 
 Having presented the approximate analytic formula for the spectrum of bound state perturbations, here we present 
 the numerical analysis to calculate the spectrum. To simplify the situation we consider the scalar spectrum with $\sigma=1$.
 
 The differential equation for the original massless scalar field $\Phi$, related to the canonically normalized field $Z$ 
 via $\Phi = Z/r$, is 
 \ba
 x (x-1) \frac{d^2 \Phi}{ dx^2}+ (2 x-1) \frac{d \Phi}{ dx} + \Big( \frac{4 \omega^2 x^3} {x-1} -\ell (\ell+1)  \Big) \Phi  =0 \, ,
 \ea
 in which we have set $r=2x$ for simplicity, so $0\leq x \leq1$ with $x=1$ being the horizon.
 
 The solution near the horizon goes like $e^{i \omega r_*} \sim (x-1)^{2 i \omega}$. Pulling out this factor from $\Phi$ and 
 defining the  new wave function via 
 \ba
 \label{Phi-psi}
 \Phi(x) \equiv (1-x)^{2 i\omega} e^{2 i \omega x} \psi(x) \, ,
 \ea 
 the corresponding equation for $\psi$ becomes
 \ba
 \label{psi-eq}
 (x^2-x) \frac{d^2 \psi}{ dx^2} + \big(4 i \omega x^2 + 2 x -1 \big) \frac{d \psi}{ dx} + \Big(4 i \omega x - \ell (\ell+1) \Big) \psi=0 \, .
 \ea
 We demand that the wave function $\psi$ to be regular everywhere in the region $0\leq x \leq1$, including at the center $x=0$.  Let us consider  the series solution for $\psi$  expanding at $x=0$, 
 \ba
 \label{series}
 \psi(x) = \sum_{m=0} c_m x^m \, .
 \ea
 The regularity of the solution at $r=0$ requires that $c_0$ to be  finite. 
 
 Plugging this series expansion into Eq. (\ref{psi-eq}) yields the following three-term recurrence relation
 \ba
 \label{recursion}
 m^2 c_m = \big[ m (m-1) -  \ell (\ell+1) \big] c_{m-1} + 4 i \omega (m-1) c_{m-2}\, , \quad  (m\ge 2)  \, ,
 \ea
with the additional condition
\ba
\label{c1-cond}
c_1= -  \ell (\ell+1) c_0 \, .
\ea
For each $m \ge 2$,  from the recursion relation Eq. (\ref{recursion}) and the condition Eq. (\ref{c1-cond}), one obtains $c_{m}$ as a polynomials of  $i \omega$ with an overall normalization proportional to $c_{0}$.

One may compare the above three-term recurrence relation with the corresponding  three-term recursion relation which have been obtained for the QNM perturbation of the exterior region \cite{Leaver:1985ax, Nollert:1993zz}. 
Demanding that $\psi$ to be regular at $x=1$ requires $\sum_{m=0} c_m$ to exist and to be finite. This is the criteria we impose to fix the spectrum of perturbations, as was done to obtain the spectra of QNMs
for the exterior region \cite{Leaver:1985ax, Nollert:1993zz}.
 
We could not find compact analytic expression for $c_m$ (i.e. the polynomial form of $c_m$ with specified coefficients  for arbitrary value of $m$) based on  the three-term recurrence relation Eq. (\ref{recursion}). However, looking at the structure of the recursion relation Eq. (\ref{recursion}) one can check that if for some special values 
of $\omega$ we manage to have $c_m=0$ for large $m$, then the successive terms $c_{m+1}, c_{m+2},...$ are suppressed  and are approximately related to $c_{m-1}$ like $c_{m+1}, c_{m+2} \sim c_{m-1}/m^2$. This suggests a rapid convergence for  $\sum_{m} c_m$, ensuring the regularity of $\psi(x=1)$. 

 We have employed the above criteria, i.e. looking for the roots of $c(m)=0$ for $m \gg1$,  to determine the spectrum of perturbations. Interestingly, we have found that our numerical values are in excellent  agreements with the numerical results 
 based on the Leaver's continued fraction technique.  In  table 1 we have presented our numerical results for  $\ell=1$ based on the method outlined above  and the corresponding values obtained from the  Leaver's continued fraction method
up to $m=60$. We see the agreement between them is excellent.   

\begin{figure}[h]
\begin{center}
	\includegraphics[scale=0.5]{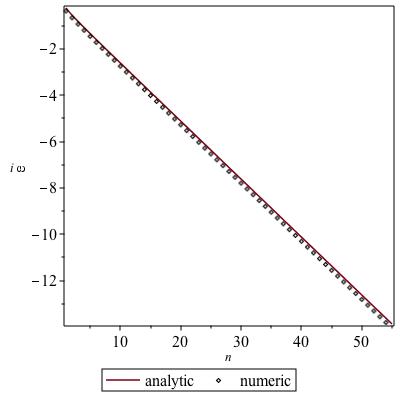} 
	\hspace{1cm}
	\end{center}
\caption{ Numerical values of $i \omega_{(n)}$. The solid curve is  from the analytic formula Eq. (\ref{spectrum}) while 
the dots are the results from our numerical approach based on the roots of $c_m=0$ for $m\gg1$.  }
\label{omega-plot}
\end{figure}

In addition, in Fig. \ref{omega-plot} we have compared the predictions of our analytic formula Eq. (\ref{spectrum}) with our 
numerical results.  A close inspection shows that while there are disagreements  for low lying spectra between these two results, but the agreement improves reasonably for high overtones, $n \gg 1$. This is not surprising;  because of the approximations employed to obtain Eq. (\ref{spectrum}), we expect it to agree with the numerical values for $n \gg 1$. As can be seen from this figure both results predict the scaling $\omega_{(n)} \simeq -\frac{i}{4} \big( n+ \frac{1}{2} \big) $ for $n \gg 1$. 
 
 \vspace{1.5 cm}

\centerline{
\begin{tabular}{|lc|c|r|}  \hline
 &our numerical method &  continued fraction method   \\  \hline
&-0.3645065834 & -0.3645065834\\  \hline
&-0.6547830675& -0.6547830672\\ \hline
&-0.9282778880&  -0.9282778882 \\ \hline
& -1.194538523&  -1.194538523\\ \hline
&  -1.456816424& -1.456816424 \\ \hline
& -1.716591508& -1.716591509\\ \hline
&  -1.974656689&  -1.974656689     \\ \hline
& -2.231484264&  -2.231484264 \\ \hline
& -2.487377381&   -2.487377382\\  \hline
\end{tabular}  }
 \vspace{1cm}

\noindent {{\bf Table 1.} In this table $i \omega$ are presented. Left: our numerical method based on the roots of $c_m=0$
for $m\gg 1$ in the series expansion Eq. (\ref{series}).  Right: Leaver's continued fraction method for the three term recursion relation Eq. (\ref{recursion}) up to $m=60$.  }

\vspace{0.5 cm}

\section{Discussions}

 In this work we have studied the  perturbations inside the   Schwarzschild black hole. There have been many works to determine the spectrum of QNM perturbations in the exterior region. The conventional view is that once the incoming QNMs cross the horizon, they are not observable as the interior of the black hole is causally disconnected from the exterior region. However, it is an interesting question to study the perturbations inside the black hole. 

 The interior of the black hole is very similar to a cosmological background with anisotropic scalings. For example, close to singularity, the spacetime looks like a Kasner background. As a result, one expects that the structure of the perturbation equation and the evolution of the perturbation to be very different than the exterior region. More specifically, the effective potential in Regge-Wheeler equation is very different than the corresponding potential in the exterior region. The structure of the new effective potential suggests the existence of the bound states with pure imaginary $\omega$. As such,  the profiles of these perturbations fall off exponentially on the boundary and are inaccessible to an outside observer.
We have calculated the spectrum of these bound states both analytically and numerically.  
 
 We have shown that our approximate analytic results are in very good agreement with the numerical results for high overtone modes $n\gg1$. We have shown that 
for $n\gg1$ the spectra of our bound states match very well with the imaginary part of the QNM spectra calculated in previous literature.  In addition, we have presented a simple method to calculate the spectra of the bound state numerically. This is based on finding the roots of the  coefficients of the three-term recurrence relation $c_m$ for large enough $m$. We have checked that the results obtained by this method are in excellent agreement obtained by the conventional continued fraction technique. 

While the spectra of our bound state are in very good agreement with the imaginary parts of the exterior QNM perturbations, however a physical link between these two solutions is missing. The bound state are not true propagating mode which is a consequence of $\omega_{(n)}$ being pure imaginary with no real components. On the other hand, the incoming QNMs are propagating wave  having a real spectrum in addition to the imaginary components. It is an interesting  question to provide a physical link between our bound state solution and the exterior QNM perturbations.

\vspace{0.7cm}

{\bf Acknowledgments:}  I am grateful to Bahram Mashhoon and Misao Sasaki for many insightful discussions and comments.  We also thank J. Abedi, M. A. Gorji  and  G. Jafari  for useful discussions.

 \appendix 
 \section{The bound state solution}  
 \label{app1}

Here we present the details of the bound state solution with $\omega^2 <0$. Starting with the approximation Eq. (\ref{V-app})  for the effective potential and defining the new variable 
\ba
y\equiv  \frac{1+ \tanh(\frac{r_*}{4})}{2 \tanh(\frac{r_*}{4})}
\ea
the wave equation is transformed into
\ba
y (y-1) \frac{d^2 Z}{dy^2}  + (2 y-1)  \frac{d Z}{dy} + \Big[ \frac{4 \omega^2}{y (y-1)} + \frac{\alpha }{4} +  \frac{\beta }{8}  \frac{2 y-1}{y-1} \Big] Z =0 \, .
\ea
Further, let us define the new variable $\rho(y)$ via
\ba
Z= y^q (y-1)^m \rho(y) \, ,
\ea
then the above differential equation is cast into
\ba
 y (y-1)  \frac{d^2 \rho}{dy^2} &+& \Big[2(1+ m+ q) y -1-2q   \Big]  \frac{d \rho}{dy} + \Big[ \frac{ (y-1)}{y}(q(q-1) + m (m-1)) + 2 m q +   \nonumber\\
&+& (2y-1) (\frac{q}{y} +  \frac{m}{y-1}) +\frac{4 \omega^2}{y (y-1)}+ \frac{\alpha }{4} +  \frac{\beta }{8}  \frac{2 y-1}{y-1}
\Big] \rho =0 \, .
\ea
We can choose the parameters $q$ and $m$ in such a way to eliminate the terms $1/y$ and $1/(y-1)$ in the terms containing 
$\rho$ in the last big bracket. This fixes $m$ and $q$ to be
\ba
m= \pm \sqrt{- 4 \omega^2 -\frac{\beta}{8}} \, ,
\ea 
and
\ba
q=\pm 2 i \omega \, .
\ea
As we shall see later on, we do not need to keep both $\pm$ branches of $m$ as both play equivalent roles in the following  analysis. Therefore, from now on,  we choose the positive branch of the above solution for $m$. 

With the above values of $m$ and $q$, the differential equation for $\rho(y)$ simplifies to
\ba
y (y-1)  \frac{d^2 \rho}{dy^2} &+& \Big[2(1+ m+ q) y -1-2q   \Big]  \frac{d \rho}{dy} + \Big[ (m+ q + \frac{1}{2})^2  + \frac{\sigma -1}{4} \Big] \rho =0  \, .
\ea
This equation has the form of differential equation for the hypergeometric function ${_2}F_1(a, \hat b, c; y)$ satisfying
\ba
\label{diff-hyper}
y (1-y) \, {_2}F_1(a, \hat b, c; y)'' + \left[c- (a+\hat b+1) y \right]    {_2}F_1(a, \hat b, c; y)' - a \hat b \,  {_2}F_1(a, \hat b, c; y) =0 \, ,
\ea
in which in our case 
\ba
a= m+ q + \frac{1}{2} + \frac{\sqrt{1- \sigma}}{2} , \quad \hat b=  m+ q + \frac{1}{2} - \frac{\sqrt{1- \sigma}}{2}, \quad
c= 1+ 2q = 1 \pm 4 i \omega \, .
\ea
Now using the relation \cite{Abramowitz}
\ba
 {_2}F_1(a, \hat b, c; y) = (1-y)^a  {_2}F_1(a, c-\hat b, c; \frac{y}{y-1}) \,  , 
\ea
the solution for $Z$ is obtained to be
\ba
Z= y^q (y-1)^{m-a}   {_2}F_1(a, c-\hat b, c; \frac{y}{y-1})  \, ,
\ea
which yields
\ba
\label{sol-final}
Z = C \Big( 1+ \tanh(\frac{r_*}{4}) \Big)^{\pm2 i \omega } \Big( 1- \tanh(\frac{r_*}{4}) \Big)^{\mp2 i \omega-\frac{1}{2} -\frac{\sqrt{1-\sigma}}{2} }  \Big( \tanh(\frac{r_*}{4}) \Big)^{\frac{1}{2} +\frac{\sqrt{1-\sigma}}{2}}  \nonumber\\
 \times \, {_2}F_1\left( a, b,c;   \frac{1+ \tanh(\frac{r_*}{4})}{ 1- \tanh(\frac{r_*}{4})}  \right) \, ,
\ea
with $a, b\equiv c-\hat b$ and $c$ given by
\ba
a& =& \frac{1}{2} +  \sqrt{- 4 \omega^2 -\frac{\beta}{8}}  + \frac{\sqrt{1- \sigma}}{2} \pm 2 i \omega \, ,   \nonumber\\
b &= &  \frac{1}{2} -\sqrt{- 4 \omega^2 -\frac{\beta}{8}}  + \frac{\sqrt{1- \sigma}}{2} \pm 2 i \omega  \, ,  \nonumber\\
c &= & 1\pm 4 i \omega  \, .
\ea

We see that at $r_* \rightarrow -\infty$,  $Z \sim e^{\pm i \omega r_*}$ as expected. To determine the  spectrum of perturbations, we demand that the wave function to be regular at $r_*=0$. Near 
$r_*=0$ the solution Eq. (\ref{sol-final}) is approximated by  
\ba
Z \rightarrow r_*^{\frac{1}{2} +\frac{\sqrt{1-\sigma}}{2}}  {_2}F_1\left( a, b,c; 1+ \frac{r_*}{2} \right) \,  \quad (r_* \rightarrow 0) \, .
\ea 
The singularity of the solution now depends on the singularity of ${_2}F_1\left( a, b,c; 1+ \frac{r_*}{2} \right)$. This in turns depends on the relation between the parameters  $a, b$ and $c$. Here the situations are somewhat different for scalar, gravitational and electromagnetic perturbations with different values of $\sigma$. 

 For the scalar perturbation ($\sigma=1$), we have $c=a+b$ in which the singularity of the hypergeometric function is given by \cite{Abramowitz}
\ba
\label{scalar-Gamma}
{_2}F_1\left( a, b,a+b; 1+ \frac{r_*}{2} \right) \rightarrow \frac{\Gamma(a+b)}{\Gamma(a) \Gamma(b)} \ln (r_*)  \, , \quad (r_* \rightarrow 0) \, .
\ea
On the other hand, for the gravitational perturbations $\sigma=-3$, we have $c=a+b-2$, and the  singularity of the hypergeometric function is given by \cite{Abramowitz}
\ba
\label{tensor-Gamma}
{_2}F_1\left( a, b,a+b-2; 1+ \frac{r_*}{2} \right) \rightarrow \frac{\Gamma(a+b-2)}{\Gamma(a) \Gamma(b) r_*^2}  (d_1+ d_2 r_*)
- \frac{ d_3 \Gamma(a+b-2)}{\Gamma(a-2) \Gamma(b-2)} \ln(r_*)  \, ,
\ea
in which $d_1, d_2$ and $d_3$ are constants. 

To impose the regularity of the wave function, we should work with the original field from which the canonically normalized 
field $Z$ is derived. Note that the auxiliary field $Z$ is defined such that the wave equation takes the form of the Schrodinger equation as given in Eq. (\ref{eq-tort}).
For the scalar perturbation, the physical (original ) field $\Phi$ is related to the auxiliary  field $Z$ via $\Phi = Z/r$.  Noting that  $r_* \sim \sqrt r$ for $r\rightarrow 0$,  and demanding that $\Phi$ to be regular at $r_*=0$, requires that the coefficient of the logarithmic 
term in Eq. (\ref{scalar-Gamma}) to vanish. However,  the $\Gamma$ function has no root so the only way to avoid the singularity is to assume that the denominators in Eqs. (\ref{scalar-Gamma})  to diverge. This happens only when either $a$ or $b$ are non-positive integers, $-n, n\ge 0$. This is the known criteria for the hypergeometric function to truncate to polynomials.

For the gravitational perturbations ($\sigma=-3$) the physical wave function, $h$, which appears in the metric perturbations 
is related to $Z$ via $h = r Z$ \cite{Chandrasekhar:1985kt}. The  axial metric perturbations which we are interested here  goes like $h \sim r$ for $r\rightarrow 0$. This in turn requires that $Z \sim r_*^{0}$. Therefore, to avoid singularity  we again need that the $\Gamma$ function in the denominator of the most singular term ($r_*^{-2}$)  in Eq. (\ref{tensor-Gamma}) to diverge. As in the case of scalar field, this requires that either $a$ or $b$ to be non-positive integers, $-n, n \ge 0$.  This conclusion also applies to electromagnetic field perturbations with $\sigma=1$. 

In summary, demanding that either $a$ or $b$ are equal to $-n$, determines the spectrum of perturbations to be  
\ba
\omega_{(n)} = \pm \frac{i}{4} \Big( n+ \frac{1}{2} +  \frac{ \sqrt{1-\sigma} }{2} - \frac{ \frac{1}{e}(\ell +\frac{1}{2})^2 +\frac{4 \sigma - e\sigma -1}{4 e}    }{n+ \frac{1}{2} +  \frac{ \sqrt{1-\sigma} }{2}}
\Big) \, .
\ea
Now, demanding that the wavefunction of the bound state $Z \propto e^{i \omega r_*}$ to fall off for  $r_* \rightarrow -\infty$, then we choose the negative branch of $\omega_{(n)}$ above.  Of course, at this level, i.e. no reference to quantum nature of the spectrum, the choice of asymptotic profile is a matter of convention.  If we choose the asymptotic form $Z \propto e^{- i \omega r_*}$, then the bound state  corresponds to the positive branch of $\omega_{(n)}$.

{}

\end{document}